\documentstyle[epsf,bibnorm]{lamuphys}
\makeatletter
\let\chapter\hid@chapter
\makeatother
\newcommand{\beq}{\begin{equation}}
\newcommand{\eeq}{\end{equation}}
\newcommand{\beqa}{\begin{eqnarray}}
\newcommand{\eeqa}{\end{eqnarray}}
\newcommand{\ba}{\begin{array}}
\newcommand{\ea}{\end{array}}
\begin{document}
\pagenumbering{arabic}

\title{Mean--Field and Nonlinear Dynamics in Many--Body Quantum Systems}

\author{V.R.\,Manfredi\inst{1,2} and L. Salasnich\inst{2,3}}

\institute{Dipartimento di Fisica ``G. Galilei'', Universit\`a di Padova, \\
Via Marzolo 8, I--35131 Padova, Italy
\and
Istituto Nazionale di Fisica Nucleare, Sezione di Padova, \\
Via Marzolo 8, I--35131 Padova, Italy
\and
Istituto Nazionale per la Fisica della Materia, Unit\`a di Milano, \\
Dipartimento di Fisica, Universit\`a di Milano, \\
Via Celoria 16, I--20133 Milano, Italy}

\maketitle

\begin{abstract}  
In this paper we discuss in detail the nonlinear equations of 
the mean--field approximation and their connection to the 
exact many--body Schr\"odinger equation. 
Then we analyze the mean--field approach 
and the nonlinear dynamics of a trapped condensate 
of weakly--interacting bosons. 
\end{abstract}

\section{Introduction}

In the past few years many authors, working in different fields, have 
shown great interest in the so--called {\it quantum chaos} or 
{\it quantum chaology}, i.e. the signature in the quantal systems 
of the chaotic properties of the corresponding ($\hbar \to 0$) 
semiclassical Hamiltonian [1--4]. Incidentally, as stressed by Berry [5], 
"the semiclassical limit $\hbar \to 0$ and the long time limit 
$t\to\infty$ are not interchangeable -- 
the origin of the $(\hbar , t^{-1})$ plane is mightily singular".   
\par
The subject is very wide but, for reasons of space, we 
focus our attention only on the connection between the 
mean--field approximation and the onset of chaos. 
For a quantum system with discrete spectrum, dynamical chaos is possible 
only as a transient with lifetime $t_H$, the so--called Heisenberg time, 
which scales as the number of degrees of freedom. 
Because $t_H$ can be very long for a many--body system, we suggest 
that the {\it transient chaotic dynamics} of quantum states and 
the related observables can be experimentally measured. 
Moreover, when the mean--field 
theory is a good approximation of the exact many--body problem, 
one can use the nonlinear mean--field equations to estimate 
the transient chaotic behaviour of the many--body system. 
As a specific example, we consider the dynamics of a trapped 
weakly--interacting Bose--Einstein condensate. 

\section{Variational principle and mean--field approximation} 

Let us consider a $N$--body quantum system with 
Hamiltonian ${\hat H}$. The exact time--dependent Schr\"odinger equation 
can be obtained by imposing the quantum last action principle on 
the Dirac action 
\beq 
S= \int dt <\psi (t) | i\hbar 
{\partial \over \partial t} - {\hat H} |\psi (t) > \; ,
\eeq
where $\psi$ is the many--body wavefunction of the system. Looking 
for stationary points of $S$ with respect to variation of the conjugate 
wavefunction $\psi^*$ gives 
\beq
i\hbar {\partial \over \partial t}\psi = {\hat H}\psi \; .
\eeq 
As is well known, it is usually impossible to obtain the exact solution 
of the many--body Schr\"odinger equation and some approximation must be used. 
\par 
In the mean--field approximation the total wavefunction 
is assumed to be composed of independent particles, i.e. it can be 
written as a product of single--particle wavefunctions $\phi_j$. 
In the case of identical fermions, $\psi$ must be antisymmetrized [6]. 
By looking for stationary action with respect to variation of a 
particular single--particle conjugate wavefunction $\phi_j^*$ one finds 
a time--dependent Hartree--Fock equation for each $\phi_j$: 
\beq
i\hbar {\partial \over \partial t}\phi_j = {\delta \over \delta \phi_j^*} 
<\psi | {\hat H}| \psi > = {\hat h} \phi_j \; ,
\eeq
where ${\hat h}$ is a one--body operator. 
The main point is that, in general, 
the one--body operator ${\hat h}$ is nonlinear. Thus 
the Hartree--Fock equations are non--linear (integro--)differential 
equations. These equations can give rise, in some cases, 
to chaotic behaviour (dynamical chaos) of the mean--field wavefunction. 

\section{Mean--Field Approximation and Chaos} 

In the mean--field approximation the 
mathematical origin of {\it dynamical chaos} resides in the nonlinearity 
of the Hartree--Fock equations. These equations provide an approximate 
description, the best independent--particle description, which 
describes, for a certain time interval, the very complicated 
evolution of the true many--body system. 
Two questions then arise: \\
1) Does this chaotic behavior persist in time? \\
2) What is the best physical situation to observe this kind of nonlinearity? 
\par
To answer the first question, 
it should be stressed that quantum systems evolve according 
to a linear equation and this is an important feature which makes them 
different from classical systems. Since the Schr\"odinger equation 
is linear, so is any of its projections. Its time
evolution follows the classical one, including chaotic
behaviour, up to $t_H$. 
After that, in contrast to the classical dynamics, 
we get localization (dynamical localization). The 
Liouville equation, on the other hand, is linear in 
classical and quantum mechanics. However, for bound 
systems, the quantum evolution operator has a purely 
discrete spectrum (therefore no long--term chaotic behaviour). 
By contrast, the classical evolution operator 
(Liouville operator) has a continuous spectrum 
(implying and allowing chaos). 
This means that persistent chaotic behaviour in the evolution 
of the states and observables is not possible. 
Loosely speaking, chaotic behaviour is possible in quantum mechanics only 
as a transient with lifetime $t_H$ [7,8]. 
\par 
The Heisenberg time, or break time, can be estimated from the Heisenberg 
indetermination principle and reads  
\beq 
t_H \simeq {\hbar \over \Delta E} \; ,
\eeq 
where $\Delta E$ is the mean energy level spacing and, 
according to the Thomas-Fermi rule, $\Delta E \propto \hbar^N$, 
where N is the number of degrees of freedom, i.e. the dimension of the
configuration space. So, as $\hbar \rightarrow 0$, the Heisenberg time 
diverges as 
\beq 
t_H \sim \hbar^{1-N} \; ,
\eeq
and it does so faster, the higher $N$ is [9]. 
We observe that the limitation to persistent chaotic dynamics 
in quantum systems does not apply if the spectrum of the Hamiltonian 
operator ${\hat H}$ is continuous. 
\par 
Concerning the second question, it is useful 
to remember that, in the thermodynamic limit, 
i.e. when the number $N$ of particles tends to 
infinity at constant density, the spectrum is, in general, continuous 
and true chaotic phenomena are not excluded [10]. 
\par 
We have seen that the Heisenberg time $t_H$ is very large for 
systems with many particles. This fact suggests 
that the {\it transient chaotic dynamics} of quantum states 
and observables can be experimentally observed in many--body quantum systems. 
Moreover, when the mean--field 
theory is a good approximation of the exact many--body problem, 
one can use the nonlinear mean--field equations to estimate 
the properties of the transient chaotic dynamics. 

\section{Nonlinear dynamics of a Bose condensate}

In this section we discuss the mean--field approximation 
and the nonlinear dynamics for a system of trapped weakly--interacting 
bosons in the same quantum state, i.e. a Bose--Einstein condensate [11]. 
In this case the Hartree--Fock equations reduce to only one equation, 
the Gross-Pitaevskii equation, which describes the dynamics 
of the condensate [12]. Nowadays, this equation is intensively studied 
because of the recent experimental achievement of Bose--Einstein 
condensation for atomic gasses in magnetic traps at very low temperatures 
(about $10^{-7}$ Kelvin) [13]. 
\par
The Hamiltonian operator of a system of $N$ identical 
bosons of mass $m$ is given by 
\beq
{\hat H}=\sum_{i=1}^N \Big( -{\hbar^2\over 2 m} \nabla_i^2 
+ V_0({\bf r}_i) \Big) + 
{1\over 2} \sum_{ij=1}^N V({\bf r}_i,{\bf r}_j) \; ,
\eeq 
where $V_0({\bf r})$ is an external potential and $V({\bf r},{\bf r}')$  
is the interaction potential. 
In the mean-field approximation the totally symmetric 
many--particle wavefunction of the Bose--Einstein condensate reads 
\beq 
\psi({\bf r}_1,...,{\bf r}_N,t) = \phi({\bf r}_1,t) ... 
\phi({\bf r}_N,t) \; , 
\eeq
where $\phi ({\bf r},t)$ is the single particle wavefunction. 
By using the quantum variational principle for the Dirac action 
we get the equation 
\beq
i\hbar {\partial \over \partial t}\phi ({\bf r},t)= 
\Big[ -{\hbar^2\over 2m} \nabla^2 
+ V_0({\bf r}) + (N-1) 
\int d^3{\bf r}' V({\bf r},{\bf r}') |\phi ({\bf r}',t)|^2 
\Big] \phi ({\bf r},t)  \; , 
\eeq
which is an integro--differential nonlinear Schr\"odinger equation. 
If the bosons are weakly interacting, it is possible to substitute 
the true interaction with a pseudo--potential 
$V({\bf r},{\bf r}') = g \delta^3 ({\bf r}-{\bf r}')$, 
where $g={4\pi \hbar^2 a_s/m}$ is the scattering amplitude and $a_s$ 
the scattering length. In this way we obtain 
the so--called Gross--Pitaevskii (GP) equation 
\beq
i\hbar {\partial \over \partial t}\phi ({\bf r},t)= 
\Big[ -{\hbar^2\over 2m} \nabla^2 
+ V_0({\bf r}) + g(N-1) |\phi ({\bf r},t)|^2 \Big] \phi ({\bf r},t)  \; . 
\eeq 
We now consider a triaxially asymmetric harmonic trapping potential of
the form 
\begin{equation}
V\left(\vec{r}\right)=
{1\over 2}m\omega_0^2\left(\lambda_1^2 x^2+
\lambda_2^2 y^2+\lambda_3^2 z^2\right)\;,
\end{equation}
where $\lambda_i$ ($i=1,2,3$) are adimensional constants proportional
to the spring constants of the potential along the three axes. 
\par 
It has been shown, using a hydrodynamical approach [14], 
that in the strong coupling limit the GP equation has exact 
solutions which satisfy a set of 
ordinary differential equations given by 
\begin{equation}
\frac{d^2}{d\tau^2}\sigma_i+\lambda_i^2\sigma_i= 
\frac{\tilde{g}}{\sigma_i\sigma_1\sigma_2\sigma_3}\;, \;\;\;\;\;\; 
i=1,2,3 \; .
\end{equation} 
These nonlinearly coupled ordinary differential equations describe 
the time evolution of the widths $\sigma$ 
of the condensate wavefunction $\psi$ along each direction 
\footnote{The same equations can be obtained by minimizing 
the Dirac action $S$ with a trial mean--field wavefunction 
$\psi({\bf r}_1,...,{\bf r}_N,t) = \phi({\bf r}_1,t)...\phi({\bf r}_N,t)$, 
where 
$$ 
\phi\left({\bf r},t\right)= 
\Big( {1\over \pi^3 a_0^6 {\sigma}_1^2(t) {\sigma}_2^2(t) 
{\sigma}_3^2(t)} \Big)^{1/4} 
\prod_{i=1,2,3} 
\exp\left\{-\frac{x_i^2}{2a_0^2 {\sigma}_i^2(t)}
+i \beta_i(t) x_i^2 \right\}\;,
$$ 
with $(x_1,x_2,x_3)\equiv(x,y,z)$. $\sigma_i$ and $\beta_i$ are 
the time-dependent variational parameters [15].}. 
Here, we have defined the coupling constant 
$\tilde{g}=(2/\pi)^{1/2}(N-1)a_s/a_0$, proportional to the
condensate number $N$ and the scattering length $a_s$. 
Note that $\tau=\omega_0 t$ and $a_0=(\hbar/m\omega_0)^{1/2}$ 
is the harmonic oscillator length. 
\par 
The three differential equations correspond to the classical 
equations of motion for a particle with coordinates $\sigma_i$ and
Hamiltonian 
\beq
H = \frac{1}{2}\left(\dot{\sigma}_1^2+\dot{\sigma}_2^2+
\dot{\sigma}_3^2\right) + 
\frac{1}{2}\left(\lambda_1^2\sigma_1^2+\lambda_2^2\sigma_2^2
+\lambda_3^2\sigma_3^2\right) 
+\tilde{g}\frac{1}{\sigma_1\sigma_2\sigma_3} \; . 
\eeq 
For $\tilde{g}\neq 0$ this Hamiltonian is nonintegrable 
and thus generic. As is well known, integrable 
systems are rather exceptional in the sense that they are typically 
isolated points in the functional space of the Hamiltonians and their 
measure is zero in this space. If we choose at random a system 
in nature, the probability that the system is nonintegrable is one [16]. 
\par 
The small oscillations and the nonlinear coupling of these 
modes have been studied by Dalfovo {et al.} for $\lambda_1=\lambda_2=1$ and 
$\lambda_3=\sqrt{8}$ (axially symmetric trap) [14]. 
One of us (L.S.) has recently calculated the mode frequencies of the low
energy excitations of the condensate in the case of 
the triaxially asymmetric potential [17]. These excitations correspond to
the small oscillations of variables $\sigma$'s around the equilibrium
point, corresponding to the minimum of the effective potential
energy of $H$. The eigenfrequencies $\omega$ 
for the collective motion, in units of $\omega_0$, are found as the
solutions of the equation 
\begin{equation} 
\omega^6
-3\left(\lambda_1^2+\lambda_2^2+\lambda_3^2\right)\omega^4
+8\left(\lambda_1^2\lambda_2^2+\lambda_1^2\lambda_3^2
+\lambda_2^2\lambda_3^2\right)\omega^2
-20\lambda_1^2\lambda_2^2\lambda_3^2
=0\;.
\end{equation} 
\par
Near the minima of the potential the trajectories in 
the phase--space are quasi--periodic. On the contrary, 
far from the minima, the effect of the nonlinearity becomes important. 
As the KAM theorem [18] predicts, 
parts of phase space become filled with chaotic orbits, 
while in other parts the toroidal 
surfaces of the integrable system are deformed but not destroyed. 
The study of this order--chaos transition for the Bose condensate 
in the triaxially asymmetric potential is currently under 
investigation by our group. 

\section{Conclusions}

The main conclusion of this paper is that the use of mean--field 
approximation leads to nonlinear equations. As a consequence, 
in some cases, the behaviour of the wavefunctions may be chaotic. 
\par 
As a specific example, the Bose--Einstein condensation 
for weakly--interacting trapped bosons has been discussed 
in great detail.

\section*{Acknowledgments}

This work has been partially supported by the Ministero 
dell'Universit\`a e della Ricerca Scientifica e Tecnologica (MURST).
L.S. thanks INFM for support through a Research Advanced Project 
on Bose-Einstein Condensation.

\end{document}